\documentstyle{article}
\begin{document}

\begin{center}
\vspace*{1.5cm}
{\Large\bf
Quantum model for magnetic multivalued recording in
coupled multilayers} \\
\vspace*{1.0cm}
  Lei Zhou, Ning Xie, Shengyu Jin\\
\vspace*{0.2cm}
{\it T. D. Lee Physics Laboratory and Department of Physics,
Fudan University,
Shanghai 200433, P. R.  China}\\
\vspace*{1.0cm}
Ruibao Tao\\
\vspace*{0.2cm}
{\it Center for Theoretical Physics, Chinese Center of Advanced
Science and
Technology (World Laboratory) ,
 P. O. Box 8730,
 Beijing 100080, China }\\
{\it and Department of Physics, Fudan University,
 Shanghai 200433, P. R. China}\\
\vspace*{1.0cm}
\end{center}

\vspace*{1.0cm}
\noindent {\bf PACS numbers:} 75.60Ej,~~75.70Fr

\vspace*{1.0cm}
\centerline{\bf\Large Abstract}
\vspace*{1.0cm}

In this paper, we discuss the possibilities of realizing the magnetic
multi-valued (MMV) recording in a magnetic coupled multilayer. The
hysteresis loop of a double-layer system is studied analytically,
and the conditions for achieving the MMV recording are given. The
conditions are studied from different respects, and the phase diagrams
for the anisotropic parameters are given in the end.

\newpage

Many efforts have been devoted to the studies of magnetic multilayers
because there have been lots of fascinating behaviors displayed in such
systems [1-4]. One of the important applications of magnetic
multilayers in technology is that they can be used as recording media
for memory devices. In such materials, the hysteresis loop of one domain
should be rather rectangular in order that two messages can be recorded
in the ``spin-up" and ``spin-down" states, respectively. Recently,
much attention has been paid to increase the density of the recording
media. One of the proposals is to diminish the size of the domain.
However, the recording density will eventually come to a limit
following this way, so that one must try to find new approaches.
A simple idea is that:
if more messages (than two) can be recorded in one domain, the recording
density will be highly improved even though the domain's size remains the
same. This is just the idea of the magnetic multi-valued (MMV) recording
which is believed to be the next strategy of high density recording and
has attracted much attention from both experimental and theoretical
sides. The MMV recording requires that more (than two) metastable phases
which are stable enough to record messages must exist in the system;
therefore, the hysteresis loop for such material should contain more
(than one) sharp steps. Experimentally, the MMV recording was firstly
confirmed by the field modulation method on disks of bilayers [3] or
island on thin layers [4].
However, the theoretical origin is not yet clear.

More recently, a quantum theory of the coercive force [5] has been
established for magnetic systems on the basis of some previous works
[6-8]. The concept of metastable state was adopted and the magnon
excitation gap was found to be an appropriate order parameter to
monitor the stability of the metastable state and to determine the
coercive force of the magnetic systems [5]. The quantum approach
enables one to study the hysteresis behaviors of a magnetic system
from a micromagnetic view, and some interesting effects in double-film
structures had been discussed by this method [5].

The present letter is devoted to proposing a theoretical possibility
of achieving the MMV recording in magnetic multilayers with the help
of the quantum method. The main idea is to find more metastable states
in such systems. It is found that the MMV recording can be realized if
the interlayer coupling and the perpendicular anisotropic constants may
satisfy some conditions.

\vspace*{1.0cm}

In this paper, a double-layer system will be investigated analytically.
The Hamiltonian can be given by:
\begin{eqnarray}
H&=&-\frac{1}{2}\sum_{m,m'}\sum_{{\bf R},{\bf R'}}
        I_{m,m'}({\bf R},{\bf R'})
        {\bf S}_{m}({\bf R})\bullet{\bf S}_{m'}({\bf R'})
        -h\sum_{m,{\bf R}}S_{m}^{z}({\bf R})\nonumber\\
 & &   - \sum_{m}\sum_{{\bf R}}D_m(S_{m}^{z}({\bf R}))^2,
\end{eqnarray}
where the subscripts $m,m'$ are the number of layers, and
${\bf R},{\bf R'}$ are the vectors of lattices on the $x$-$y$ plane.
$I_{m,m'}({\bf R},{\bf R'})$ are the exchange parameters and only the
nearest-neighbor interaction is considered. The
single-ion anisotropy is the ``easy-axis" case ($D_m > 0$), and the
``easy-axis" is perpendicular to the film. The spins and the
anisotropies in different layers are different. It is supposed that
$S_1 > S_2$ without loosing any generality.

Following Refs. 5-6, we will introduce the local coordinates (LC)
system $\{\hat x_m, \hat y_m, \hat z_m\}$. The spin components in the LC
system will have the following relations with those in the original one:
\begin{math}
S_m^x = \cos\theta_m S_m^{x_m} + \sin\theta_m S_m^{z_m},
S_m^y = S_m^{y_m}({\bf r}),
S_m^z = \cos\theta_m S_m^{z_m} - \sin\theta_m S_m^{x_m}.
\end{math}
In order to study the ground state properties and the low-lying
spin-wave excitations, one can apply the usual spin-Bose
transformation such as Holstein-Primakoff (H-P) [9]
or the complete Bose transformations (CBT) [10] to the spin
operators in the LC system $\{ S_m^{x_m},S_m^{y_m},S_m^{z_m}\}$. In a
harmonic approximation, the H-P transformation and the CBT yield the
same results. Then, after the LC transformation and the Bose
transformation, the Hamiltonian becomes:
\begin{eqnarray}
H=U_0 + H_1 + H_2 +\cdots,
\end{eqnarray}
$H_2$ can be written in the momentum ${\bf k}$ space for
$x$-$y$ plane as following:
\begin{eqnarray}
H_2 &=& \sum_{m,m'}\sum_{{\bf k}} F_{m,m'}({\bf k},\theta)
         a_m^+({\bf k})a_{m'}({\bf k})\nonumber\\
    &+& \sum_{m,m'}\sum_{{\bf k}} G_{m,m'} ({\bf k}, \theta)
        [a_m^+({\bf k})a_{m'}^+(-{\bf k})+a_{m}({\bf k})
                a_{m'}(-{\bf k})],
\end{eqnarray}
in which the coefficients $F_{m,m'}$ and $G_{m,m'}$ are defined by:
\begin{eqnarray}
   F_{m,m}({\bf k},\theta)
&=&I_{m,m}ZS_m(1-\gamma_k) - D_m(S_m-\frac{1}{2})
                (\sin^2\theta_m-2\cos^2\theta_m)\nonumber\\
& &+\sum_{m'}S_{m'}I_{m,m'}\cos(\theta_m-\theta_m')
        +h\cos\theta_m,\label{eq:fm}\\
       F_{m,m'}({\bf k}, \theta)
&=&-\frac{1}{2}I_{m,m'}\sqrt{S_mS_{m'}}
        [1+\cos(\theta_m-\theta_{m'})],\\
   G_{m,m}({\bf k},\theta)
&=&-\frac{1}{4}\sqrt{2S_m(2S_m-1)} D_{m} \sin^2\theta_m,\\
   G_{m,m'}({\bf k}, \theta)
&=&\frac{1}{4}I_{m,m'}\sqrt{S_mS_{m'}}
        [1-\cos(\theta_m-\theta_{m'})].\label{eq:gm}
\end{eqnarray}
Here, $\gamma_k = (1/Z)\sum_{\delta}exp(i{\bf k}\cdot {\bf\delta})$
where the summation $\delta$ runs over the $Z$ nearest-neighbors
of a given site in the $x$-$y$ plane.

In a first order approximation, spin configuration \{$\theta_m$\} can
be obtained by minimizing the ground state energy $U_0$:
$\delta U_0 / \delta\theta_m = 0$, which yield the following equations:
\begin{eqnarray}
\sum_{m'}I_{m,m'}S_{m'}
\sin(\theta_m-\theta_m') + h\sin\theta_m
+D_m(2S_m-1)\sin\theta_m\cos\theta_m = 0 ~ m=1,2,\cdots.\label{eq:nl}
\end{eqnarray}
Equations above are just the same as the condition of $H_1=0$.
The harmonic part of Hamiltonian can be exactly diagonalized by
a generalized Bogolyubov transformation:
\begin{eqnarray}
\alpha_m^+({\bf k})~~~~&=& \sum_n U_{m,n}({\bf k}) a_n^+({\bf k})
                      +\sum_n V_{m,n}({\bf k}) a_n(-{\bf k})
                        \label{eq:bg1}\\
\alpha_m(-{\bf k})     &=& \sum_n U_{m,n}({\bf k}) a_n(-{\bf k})
                   +\sum_n V_{m,n}({\bf k}) a_n^+({\bf k}),
                        \label{eq:bg2}
\end{eqnarray}
so that we get finally:
\begin{eqnarray}
H = U_0' + \sum_{k} \epsilon_m({\bf k})
\alpha_m^+({\bf k})\alpha_m({\bf k}) + \cdots \label{eq:hd}
\end{eqnarray}
where the magnon excitation energy $\epsilon_m({\bf k})$ in
Eq. (\ref{eq:hd}) and the coefficients $(U_{m,n},V_{m,n})$ in
Eqs. (\ref{eq:bg1})-(\ref{eq:bg2}) can be obtained from
the eigenvalues and the eigenvectors of the following matrix
\begin{eqnarray}
   {\hat {\cal H}}({\bf k})=
    \left( \begin{array}{ccc}
{\hat {\cal F}}({\bf k})      &  i~ 2{\hat {\cal G}}({\bf k}) \\
i ~2{\hat {\cal G}} ({\bf k}) &   -{\hat {\cal F}} ({\bf k})\\
  \end{array} \right)\label{eq:hk}.
\end{eqnarray}
The elements of the sub-matrices ${\cal F}({\bf k})$
and ${\cal G}({\bf k})$ in matrix ${\cal H}({\bf k})$ are
$F_{m,m'}(\theta,{\bf k})$ and $G_{m,m'}(\theta, {\bf k})$
defined in Eqs. (\ref{eq:fm})-(\ref{eq:gm}), respectively [5].

Following Ref. 5, the minimum value of the magnon excitation energy
$\epsilon_m({\bf k})$ is defined as the gap:
$\Delta(h)=Min[\epsilon_m({\bf k})]$.

Eqs. (\ref{eq:nl}) may have
many solutions corresponding to various possible spin configurations.
For every solution of Eqs. (\ref{eq:nl}), one can calculate the magnon
excitation gap $\Delta(h)$ following the method described above.
According to Ref. 5, if the gap $\Delta(h)$ is positive, the state
described by such a solution is a metastable one since a variation
from this state must cost energy. However, when the gap comes to
zero even negative at a field $h_c$, such a state will no longer
be metastable and a transition from this state to another metastable
one will take place. Thus, in the case that there are many
metastable states existing in the system, the MMV recording is possible
to take place.

Eq. (\ref{eq:nl}) has two kinds of solutions:
the trivial solutions (i.e. $\theta_{m} = 0$ or,$\pi$,~~$m=1,2$)
which correspond to the aligned spin states; the nontrivial solutions
(i.e. $\theta_{m} \not= 0$ or,$\pi$,~~$m=1,2$) to the
canted spin states. Subsequently, we will discuss both the aligned spin
states and the canted spin states, and discuss which state the system will
transit to if the current state is unstable.

The following notations will be used.
$I_{m,m}({\bf R},{\bf R'})=J$,
$I_{m,m'}({\bf R},{\bf R}) = I $ and $D_m(2S_m-1)=\tilde D_m$. The
exchange interaction within a layer should be the ferromagnetic
type ($J > 0$). However, both the ferromagnetic and the
antiferromagnetic types of interlayer exchange coupling
will be discussed (i.e. $I>0$ or $I<0$ ).\\

\noindent{\bf\Large The aligned spin states:}\\

In such a system, four aligned states are possible.
They are illustrated in figure 1.
 
For A configuration, we have $\theta_1=0, \theta_2=0$.
From Eqs. (\ref{eq:fm})-(\ref{eq:gm}), we obtain
\begin{equation}
   {\hat {\cal F}}({\bf k})=
    \left( \begin{array}{cc}
IS_2 + \tilde D_1
+ h+JZS_2(1-\gamma_k)   &  -I\sqrt{S_1S_2} \\
-I\sqrt{S_1S_2}         &  IS_1 + \tilde D_2 +
                          h + JZS_1(1-\gamma_k)\\
  \end{array} \right)\label{eq:fkm}
\end{equation}
and
\begin{eqnarray}
{\hat {\cal G}}({\bf k}) = 0\label{eq:gkm}
\end{eqnarray}
According to Eqs. (\ref{eq:hd})-(\ref{eq:hk}), we find that
the excitation energy $\epsilon_m{(\bf k)}$ are
just the eigenvalues of the matrix $\hat {\cal F}({\bf k})$. Thus we
obtain:
\begin{eqnarray}
\Delta_{A}(h)= h +
\frac{1}{2} [\tilde D_1 + \tilde D_2 + IS_1 + IS_2
- \sqrt{(IS_2-IS_1+\tilde D_1 -\tilde D_2)^2 + 4I^2S_1S_2}]
\end{eqnarray}
From the discussions above, one may find that the system in
A configuration will be stable only in the case that
\begin{eqnarray}
h \ge h_c^0,
\end{eqnarray}
where
\begin{eqnarray}
h_c^0 =
\frac{1}{2} [-IS_1 - IS_2 - \tilde D_1 - \tilde D_2
+ \sqrt{(IS_2-IS_1+\tilde D_1 -\tilde D_2)^2 + 4I^2S_1S_2}].
\end{eqnarray}

It is similar for B configuration. One may find that the stable
region for B configuration is
\begin{eqnarray}
h_c^2 \le h \le h_c^1,
\end{eqnarray}
where
\begin{eqnarray}
h_c^1 = \frac{1}{2} [IS_2 - IS_1 +\tilde D_2 - \tilde D_1 +
\sqrt{(\tilde D_1+\tilde D_2-IS_1-IS_2)^2-4I^2S_1S_2}],\\
h_c^2 = \frac{1}{2} [IS_2 - IS_1 +\tilde D_2 - \tilde D_1 -
\sqrt{(\tilde D_1+\tilde D_2-IS_1-IS_2)^2-4I^2S_1S_2}].
\end{eqnarray}

Considering the symmetry between A,B states and C,D states,
it is very easy to understand that the stable region of C and D states
are $[-h_c^1, -h_c^2]$ and $(-\infty, -h_c^0]$, respectively.\\

\noindent{\bf\Large The canted spin states}\\

For every trivial solution, non-trivial solutions can be
bifurcated from them at some fields.
Around the bifurcation points, the variations of the angles
from the trivial solution should be very small.
Thus, it is reasonable to linearize the
non-linear equations (\ref{eq:nl}) to study the behaviors of the
non-trivial solutions around the bifurcation point. Taking A
configuration as an example, we have: $\sin\theta_m \sim \theta_m$.
Then around the bifurcation point, equations (\ref{eq:nl}) will
be linearized as:
\begin{eqnarray}
IS_2 (\theta_1 - \theta_2) + h\theta_1 + \tilde D_1\theta_1 = 0
\label{eq:ln1}\\
IS_1 (\theta_2 - \theta_1) + h\theta_2 + \tilde D_2\theta_2 = 0
\label{eq:ln2}
\end{eqnarray}
which can be rewritten as:
\begin{eqnarray}
\left(\begin{array}{cc}
        -\tilde D_1 - IS_2 & IS_2\\
        IS_1               & -\tilde D_2 - IS_1\\
      \end{array}\right)
\left(\begin{array}{c}
        \theta_1\\
        \theta_2\\
        \end{array}\right) = h
\left(\begin{array}{c}
        \theta_1\\
        \theta_2\\
        \end{array}\right)\label{eq:ln}.
\end{eqnarray}
The matrix in the left side of the above equation can be diagonalized
with the following two eigenvalues:
\begin{eqnarray}
\lambda_1 =
\frac{1}{2} [-IS_1 - IS_2 - \tilde D_1 - \tilde D_2
+ \sqrt{(IS_2-IS_1+\tilde D_1 -\tilde D_2)^2 + 4I^2S_1S_2}],\\
\lambda_2 =
\frac{1}{2} [-IS_1 - IS_2 - \tilde D_1 - \tilde D_2
- \sqrt{(IS_2-IS_1+\tilde D_1 -\tilde D_2)^2 + 4I^2S_1S_2}].
\end{eqnarray}
Thus, if $h=\lambda_1$ or $\lambda_2$, the linearized equations
(\ref{eq:ln1})-(\ref{eq:ln2}) or Eq. (\ref{eq:ln}) may have
non-zero solutions. If we put this solution as a ``guessed solution"
into the non-linear equations (\ref{eq:nl}), we can finally get a
non-trivial solution step by step. The non-trivial solutions
bifurcated from other trivial solutions can be studied similarly.

Noting $\lambda_1 > \lambda_2$, it is easy to understand that
the non-trivial solution can not exist when $h  > \lambda_1$.
It is interesting to find that $\lambda_1 = h_c^0$. Thus the
canted spin state can appear only in the case that the aligned
spin state is not stable.

In order to study whether the canted spin state can be stable
or not, one should study the magnon excitation gap $\Delta(h)$ in the
vicinity of the critical point $h_c^0$ for the non-trivial solution.
Since the angles $\theta_1, \theta_2$ are very small in this case,
it is reasonable to adopt a first order approximation
when calculating the magnon excitation gap. From Eqs.
(\ref{eq:fm})-(\ref{eq:gm}), the elements of
the matrix $\hat {\cal H}(h_c^0, {\bf k})|_{\bf k=0}$
in a first order approximation can be given as:
\begin{eqnarray}
F_{m,m'} = F_{m,m'}^0 + \delta F_{m,m'},\\
G_{m,m'} = G_{m,m'}^0 + \delta G_{m,m'},
\end{eqnarray}
where $F_{m,m'}^0, G_{m,m'}^0$ have been defined in Eqs.
(\ref{eq:fkm})-(\ref{eq:gkm}) by setting ${\bf k}=0$.
$\delta F_{m,m'}$ are found to be:
\begin{eqnarray}
\delta F_{1,1} = -\frac{1}{2}IS_2(\theta_1-\theta_2)^2 -
(\frac{1}{2}h_c^0 + \frac{3}{2}\tilde D_1)\theta_1^2\label{eq:df1}\\
\delta F_{2,2} = -\frac{1}{2}IS_1(\theta_1-\theta_2)^2 -
(\frac{1}{2}h_c^0 + \frac{3}{2}\tilde D_1)\theta_2^2\\
\delta F_{1,2} = \delta F_{2,1} = -
\frac{1}{4}\sqrt{I^2S_1S_2}(\theta_1-\theta_2)^2\label{eq:df3}
\end{eqnarray}
Since $\theta_1, \theta_2$ can be substituted by the solution of the
linearized equations (\ref{eq:ln1})-(\ref{eq:ln2})
in the critical point $h_c^0$ in a first order approximation,
they must have the following relation:
\begin{eqnarray}
\frac{\theta_1}{\theta_2} = \frac{IS_2}{IS_2 + \tilde D_1 + h_c^0}
= \frac{IS_2}{F_{1,1}^0}\label{eq:rlt}
\end{eqnarray}
Thus all the terms in Eqs. (\ref{eq:df1})-(\ref{eq:df3})
can be obtained after extracting
a common parameter $\theta_2^2$ through use of Eq. (\ref{eq:rlt}).
For example,
\begin{eqnarray}
\delta F_{1,1} = [-\frac{1}{2}IS_2(1-\frac{IS_2}{F_{1,1}^{0}})^2
-(\frac{h_c^0}{2}+\frac{3}{2}\tilde D_1)
(\frac{IS_2}{F_{1,1}^{0}})^2] \theta_2^2.
\end{eqnarray}
So, based on the perturbation theory, the magnon excitation gap
$\Delta(h_c^0)$ of the canted spin state at the critical point
$h_c^0$ in the first order approximation can be presented as:
\begin{eqnarray}
\Delta(h^0_c) \simeq \delta F_{1,1} + \delta F_{2,2} -
\frac{1}{F_{1,1}^0 + F_{2,2}^0}
[(\delta F_{1,1} - \delta F_{2,2})(F_{1,1}^0 - F_{2,2}^0) +
4\sqrt{I^2S_1S_2}\delta F_{1,2}].
\end{eqnarray}
$\Delta(h^0_c)/\theta_2^2$ must now be a definite value determined by
the parameters.
The non-trivial spin state which is bifurcated from the B configuration
can be studied similarly, and the magnon excitation gap $\Delta(h_c^2)$
for such a canted spin states can also be derived following the same
procedure.\\

\noindent{\bf\Large The conditions for MMV recording}\\

We have studied both the aligned spin state and the canted one. In order
to realize the MMV recording, the four aligned spin states must be
overlapping with each other. Thus, it is required that:
\begin{eqnarray}
h_c^2 < h_c^0 < h_c^1 \label{eq:cdt1}.
\end{eqnarray}
On the other hand, to be used for recording, the hysteresis loop
should be as sharp as possible. Otherwise, it may cause difficulty to
distinguish two messages. Thus, the canted spin states should not exist:
\begin{eqnarray}
\Delta(h_c^0) < 0, ~~~~~ \Delta(h_c^2)  < 0 \label{eq:cdt2}.
\end{eqnarray}
(\ref{eq:cdt1})-(\ref{eq:cdt2}) are just the conditions for realizing
the MMV recording in a double-layer structure. The conditions are the
complicated relations between the single-ion anisotropy parameters
($D_1, D_2$), the exchange interaction parameter ($I$) and the spins
($S_1$, $S_2$). We will study them from different respects.

First, we study what is the requirement for the interlayer exchange
parameter $I$ if the two magnetic layers are determined. The following
model will be investigated:

\hspace*{2.0cm} Model 1:
$S_1 = 3, S_2 = 1, D_1/D_2=2.0$

\noindent The critical fields $h_c^0, h_c^1, h_c^2$ have been shown
together as functions of $I/D_2$ in figure 2. One may find that
the exchange parameter should satisfy $I_c^1 < I < I_c^2$ for
condition (\ref{eq:cdt1}). If the exchange coupling is the ferromagnetic
case and is very strong ($I>I_c'$),
the B and C states can not be stable at all since the two magnetic layers
are unwilling to antiparallel with each other (fig. 2).
In figure 3, $\Delta(h_c^0)$ is shown with respect to $I/D_2$
in order to study the stabilities of the canted spin state bifurcated
from A configuration. One may find that: only
when the interlayer coupling is the antiferromagnetic case and is
stronger than a critical value ($|I| < |I_c^3|$), could this canted
spin states appear. If the coupling is the ferromagnetic case,
this canted spin state is not able to be metastable. $\Delta(h_c^2)$
has also been studied for model 1, and it is always negative.
In all, the exchange coupling should {\bf not} be very strong compared
to the anisotropy in order to realize the MMV recording. An example has
been shown in figure 4 where $I/D_2 = 0.5$. The
multi-step shape of the hysteresis loop can clearly observed.

However, there remains a question. At the field $h_c^0$ where
A spin configuration is no longer stable, B and D spin states are
both stable. Why the system will transit to B spin configuration instead
of D configuration (fig. 4) ? This question can be answered by studying the
magnon excitation spectrum. According to Eqs. (\ref{eq:bg1})-(\ref{eq:hk}),
one can get the concrete forms of the Bose operators $\alpha_m({\bf k})$.
We only study the lowest mode of spin wave, so that ${\bf k}=0$.
Suppose $m=1$ without losing generality, thus
\begin{eqnarray}
\epsilon_1 (h_c^0) &=& \Delta(h_c^0) = 0,\\
\alpha_1 &=& U_{1,1} a_1 + U_{1,2} a_2
\end{eqnarray}
where
\begin{eqnarray}
U_{1,1} = \frac{I\sqrt{S_1S_2}}
{\sqrt{(IS_2+h_c^0+\tilde D_1)^2 + I^2S_1S_2}}\\
U_{1,2} = \frac{IS_2+h_c^0+\tilde D_1}
{\sqrt{(IS_2+h_c^0+\tilde D_1)^2 + I^2S_1S_2}}
\end{eqnarray}

Since the excitation energy of this mode is zero, if there are
any kind of fluctuations, the Bosons at this mode must be greatly
excited without costing energy. The current spin configuration will be
completely destroyed because of the excitations. Noting $\alpha_1$ is
a linear combination of $a_1,a_2$, the quantities $|U_{1,m'}|^2$ may be
understood as the possibilities of the Bosons in the $m'$th layer to be
excited, thus they must can be considered as the possibilities of the
spins in the $m'$th layer to turn flipping. In figure 5, the two
quantities $|U_{1,m'}|^2$
are shown together as functions of $I/D_2$ for Model 1. One may find
that in the region where the MMV recording is permitted,
$|U_{1,1}|^2 \sim 0$ while $|U_{1,1}|^2 \sim 1$. Thus, at the field
$h_c^0$ where the A configuration is not able to be stable, the spins in
2nd layer are mostly likely to turn flipping while those in the 1st layer
are not likely to do so. So, in this case, the system will
transit to B configuration instead of D configuration. One may also find
that if the interlayer exchange is the ferromagnetic case and is
very strong ($I \gg D_2$), the two quantities will be close.
Thus, the two magnetic layers are willing to turn flipping together
because of the strong interlayer coupling.
The MMV recording is not able to be realized then. By the way, if there
is no coupling between the two magnetic layers ($I=0$), we find that
$|U_{1,1}|^2=0$ and $|U_{1,2}|^2=1$. This is easy to be understood. Because
the two layers are not coupled, they can be treated independently. In
this case, one may find that $h_c^0$ and $h_c^2$ are just the coercive
forces of the two layers. So, when the external field reaches $h_c^0$,
the 2nd layer will turn over while the first layer will not.

\vspace*{1.0cm}

Finally, We study what kinds of materials can be used for MMV
recording. Suppose $S_1=S_2=S,  D'_m = (2S-1)D_m / |I|S $,
the phase diagrams for $ D'_1 - D'_2$ plane are given in
ferromagnetic case (figure 6) and in antiferromagnetic case (figure 7),
respectively. The two cases are quite different. To realize the MMV
recording, the anisotropies for the two materials can not be very
close ($D_1' \sim D_2'$) if the exchange is the ferromagnetic case
(fig. 6), while there is no such restriction for
the antiferromagnetic case (fig. 7). However, a common requirement in
the two cases is still a weak interlayer exchange interaction.

If the structure is more complicated, one may easily understand that
there may be more sharp steps in the hysteresis loop. Thus even more
messages can be recorded in one domain. The extension from the
present double-layer system to a complicated one is rather
straightforward.

\noindent{\bf\large Conclusion}\\

In conclusion, we have analytically studied the hysteresis loop of
a double-layer magnetic system. We find that when the interlayer
coupling and the anisotropies of the two materials satisfy some
complicated conditions, more metastable states will
be possible to appear, and the magnetic multi-valued recording may
be realized. The conditions are discussed from different respects, and
the permitted values of the anisotropies for realizing the MMV recording
are presented.\\

\noindent{\large\bf Acknowledgments}\\

\noindent This research is supported by National Science Foundation of
China and the National Education commission under the grant for training
Ph.Ds.

\vspace*{1.0cm}
\noindent {\large\bf Captions:}

\vspace*{0.5cm}

\noindent Figure 1: Aligned spin configurations in a double-layer
magnetic system.

\vspace*{0.5cm}

\noindent Figure 2: Critical fields $h^0_c$, $h_c^1$ and $h_c^2$ as
functions of the interlayer coupling constant for the double-layer
system.

\vspace*{0.5cm}

\noindent Figure 3: $\Delta(h_c^0)/\theta_2^2$ as the function of the
interlayer coupling constant for the double-layer system.

\vspace*{0.5cm}

\noindent Figure 4: Hysteresis loop of a double-layer magnetic system
with a ferromagnetically interlayer coupling constant $I / D_2 = 0.5$.

\vspace*{0.5cm}

\noindent Figure 5: The possibilities for the two magnetic layers to turn
flipping ($|U_{1,1}|^2$, $|U_{1,2}|^2$) as functions of the
interlayer exchange parameter.

\vspace*{0.5cm}

\noindent Figure 6: Permitted values of the two anisotropies for
realizing the MMV recording in the ferromagnetic coupling case. One can
achieve the MMV recording in region I, and can not do so in region II.

\vspace*{0.5cm}

\noindent Figure 7: Permitted values of the two anisotropies for
realizing the MMV recording in the antiferromagnetic coupling case.
One can achieve the MMV recording in region I, and can not do so
in region II.

\end{document}